\documentclass[prd,aps,amsmath,%
floats,a4paper,twocolumn,%
tightenlines,notitlepage,%
superscriptaddress,%
]{revtex4}
\usepackage{epsfig}
\begin{document}
\def\lsi{\raise0.3ex\hbox{$<$\kern-0.75em\raise-1.1ex\hbox{$\sim$}}}
\def\gsi{\raise0.3ex\hbox{$>$\kern-0.75em\raise-1.1ex\hbox{$\sim$}}}
\newcommand{\lsim}{\mathop{\lsi}}
\newcommand{\gsim}{\mathop{\gsi}}
\def\tphi{\tilde{\chi}}
\def\tE{\tilde{E}}
\def\xpi{x\!+\!\hat{\imath}}
\def\xpj{x\!+\!\hat{\jmath}}
\def\xpk{x\!+\!\hat{k}}
\def\sd{\dot\varphi}


\title{Dynamics of tachyonic preheating after hybrid inflation}
\author{
E.J. Copeland\email{e.j.copeland@sussex.ac.uk}}
\affiliation{
Centre for Theoretical Physics,
University of Sussex, Falmer, Brighton BN1 9QJ,~U.~K.}
\author{
S. Pascoli\email{pascoli@sissa.it}}
\affiliation{
Scuola Internazionale Superiore di Studi Avanzati,
via Beirut 2-4, I-34014, Trieste, Italy\\ 
and INFN, Sezione di Trieste,
 I-34014, Trieste, Italy}
\author{
A. Rajantie\email{a.k.rajantie@damtp.cam.ac.uk}
}
\affiliation{
DAMTP, CMS, University of Cambridge,
Wilberforce Rd, Cambridge, CB3 0WA,~U.~K.
}

\date{March 20, 2002}

\begin{abstract}
We study 
the instability of a scalar field at the end of hybrid inflation, 
using both analytical techniques and numerical simulations.
We improve  previous studies by taking the inflaton field fully
into account, and show that the range of unstable modes depends
sensitively on the velocity of the inflaton field, and thereby on the
Hubble rate, at the end of inflation.
If topological
defects are formed, their number density is determined by the shortest
unstable wavelength.
Finally, we show that the oscillations of the inflaton field amplify
the inhomogeneities in the energy density, leading to local symmetry
restoration and faster thermalization. We believe this explains why
tachyonic preheating is so effective in transferring energy away from
the inflaton zero mode.

\vspace*{.3cm}
\noindent DAMTP-2002-1, SUSX-TH-02-002, SISSA/2/2002/EP
 \hfill
hep-ph/0202031
\end{abstract}
\maketitle


\section{Introduction}
The dynamics of reheating after inflation have been 
studied intensively~\cite{Dolgov:1982th,Abbott:1982hn}
ever since the first models of inflation were formulated. 
In many cases, this process takes place non-perturbatively
via a parametric resonance between the inflaton and matter
fields~\cite{Traschen:1990sw,Kofman:1994rk}. In this scenario, which is
known as preheating, the long-wavelength matter modes are 
exponentially amplified to a very high 
effective temperature.
This may have the consequence that
certain symmetries, which are spontaneously broken at the reheat
temperature, are temporarily restored during this non-thermal
stage~\cite{Kofman:1996fi,Tkachev:1996md}. 
When the fields thermalize, the symmetry breaks
again, leading possibly to formation of 
topological defects~\cite{Tkachev:1998dc}.
In models of electroweak scale inflation, the period of non-thermal
symmetry restoration may also explain the baryon asymmetry of the
universe~\cite{Krauss:1999ng,Garcia-Bellido:1999sv,Rajantie:2001nj}.

However, it was argued more recently by
Felder {\it et al.}~\cite{Felder:2001hj} 
that in hybrid inflationary models, this parametric resonance often
fails to
take place. In these models, inflation ends because a scalar field
becomes unstable and its long-wavelength modes grow while the
inflaton field rolls down the potential.
The dynamics of this ``tachyonic'' or spinodal instability have been studied
by many authors~\cite{Boyanovsky:1997rw,Garcia-Bellido:1998wm,Felder:2001hj,%
Copeland:2001qw,Asaka:2001ez,%
Felder:2001kt,Garcia-Bellido:2001cb,DeMelo:2001nr}, using a range of
different methods.

Naively, the inflaton field would be expected
to start oscillating once it reaches the minimum of the potential. What
Felder {\it et al.}~\cite{Felder:2001hj} observed was 
that the oscillations die
away almost immediately, and coined the term ``tachyonic preheating''
to describe this phenomenon. 
This damping rules out the possibility of ordinary preheating, but
the tachyonic instability itself still can lead to many similar
effects, such as
generation of baryon
asymmetry~\cite{Copeland:2001qw,Garcia-Bellido:2001cb}, defect
formation~\cite{Felder:2001hj,Asaka:2001ez} and particle 
production~\cite{Garcia-Bellido:2001cb,DeMelo:2001nr}.
Nevertheless, the physical origin of the damping has remained unclear
until now.

In this paper, we use both detailed analytical calculations and
non-perturbative numerical simulations to study the dynamics of
tachyonic preheating after hybrid inflation. Analytical results 
show that the upper
limit, $k_*$, of the range of momenta amplified during the instability
has a simple
dependence on the velocity $\sd$ of the inflaton field at the instability
point, 
$k_*\approx
(2mg\sd)^{1/3}$~\cite{Garcia-Bellido:1998wm,Copeland:2001qw}. 
This is essentially
equivalent to calculations of the number of topological defects
formed in a phase transition~\cite{Zurek:1985qw,Karra:1997it,Bowick:1998kd}. 

We confirm this result in our numerical simulations, in which we
include also the inflaton as a dynamical field, by measuring the power
spectrum directly and also by measuring the number of topological
defects formed in the transition. Furthermore, we study the later
evolution of the system, and show that because the potential in
hybrid models is extremely flat above the critical value, the
oscillations of the inflaton field form
roughly spherical 
``hot spots'', inside which the energy density
is very high and symmetry restoration can take place locally. 
Depending on the values of the couplings, these hot spots may appear
already during the first oscillation of the inflaton field.
We believe this explains the strong dissipation reported in 
Ref.~\cite{Felder:2001hj}, where the inflaton field was observed to
settle down in its minimum after only  few oscillations.

We carry out most of our simulations in two dimensions,
because that allows us to treat a wider range of length scales in a
single simulation, but in order to check that our results are
not specific of two dimensions, we repeat some of the simulations in 
three dimensions. 
The breaking of global  symmetries 
leads to the existence of massless 
Goldstone bosons, which are instead not present
in the case of the breaking of local gauge symmetries.
We recall that there is no experimental
evidence for the existence of such 
scalars whose couplings to the particles
of the Standard Model of Particle Physics
are severely constrained by the present data.
Therefore we study also the case in which
the broken symmetry is a local gauge invariance.

The structure of the paper is the following: 
We start by reviewing the basic features of hybrid inflationary models
in Section~\ref{sect:hybrid}.
In Section~\ref{sect:instability}, we study the instability of the
scalar field analytically both in the approximation that the mass term
changes instantaneously and in the more realistic case of hybrid
inflation.
We present the details of our two-dimensional
lattice simulations, together with the
results for the power spectrum in Section~\ref{sec:simu}, and the
results for the defect density in Section~\ref{sect:defects}. In
Section~\ref{sect:locres}, we show that the oscillations of the
inflaton field lead to ``hot spots'' with local symmetry restoration.
In Section~\ref{sect:3dsimu}, we 
present results of
three-dimensional simulations showing that our findings apply to that
case as well.

\section{Hybrid inflation}
\label{sect:hybrid}
In models of hybrid inflation~\cite{Linde:1994cn}, 
the inflaton field $\varphi$ is coupled to
another scalar field $\chi$, which becomes unstable at a certain critical
value of $\varphi=\varphi_c$. 
This causes the slow roll conditions to break
down, and inflation ends. The simplest realization of this idea has
the potential
\begin{equation}
\label{equ:hybrid_pot}
V(\varphi,\chi)=\frac{1}{2}m_\varphi^2\varphi^2+\frac{1}{2}g^2\varphi^2\chi^2
+\frac{1}{4}\lambda\left(\chi^2-v^2\right)^2,
\end{equation}
where $\varphi$ and $\chi$ are real scalars.
During inflation, the inflaton has a large value,
$\varphi\gg\varphi_c=m/g$, where we have defined $m\equiv\lambda^{1/2}v$.
Therefore the effective mass term of the
$\chi$ field, $m_\chi^2(\varphi)=-m^2+g^2\varphi^2$, is positive
and the symmetry $\chi\rightarrow-\chi$ is restored.
Eventually, $\varphi$ reaches $\varphi_c$, and
$m_\chi^2(\varphi)$ becomes negative, implying that the $\chi$ field
becomes unstable, or tachyonic.

The COBE measurements require
that at $\varphi=\varphi_{\rm COBE}$, when the fluctuations in the
cosmic microwave background left the horizon,
\begin{equation}
\frac{V(\varphi_{\rm COBE})^{3/2}}{M_p^3V'(\varphi_{\rm COBE})}\approx
5.3\times 10^{-4},
\end{equation}
where $M_p=(8\pi G)^{-1/2}\approx 2.4\times 10^{18} {\rm GeV}$ is the
reduced Planck mass.
Assuming that the value of the potential and its derivative do
not change significantly between $\varphi_{\rm COBE}$ and $\varphi_c$,
we find
\begin{equation}
V'(\varphi_{\rm COBE})\approx m_\varphi^2\varphi_c=\frac{mm_\varphi^2}{g},
\end{equation}
and $V(\varphi_{\rm COBE})\approx V_0\equiv m^4/4\lambda$. 
Consequently,
\begin{equation}
m_\varphi^2\approx \frac{1}{5.3\times 10^{-4}}
\frac{g}{8\lambda^{3/2}}\frac{m^5}{M_p^3},
\end{equation}
which is typically much less than $m^2$, and therefore we neglect it
in the study of the dynamics of the system.
If the slow roll 
condition $V'(\varphi)M_P\ll V(\varphi)$ is satisfied, we have
\begin{equation}
\label{equ:slowroll}
\sd=\frac{V'}{3H}=\frac{V'M_P}{\sqrt{3V}}
\approx
\frac{1}{5.3\times 10^{-4}}\frac{V_0}{\sqrt{3}M_P^2}\approx (60 H)^2,
\end{equation}
where $H$ is the Hubble rate at the end of inflation.
Even in GUT scale inflation, $\sd$ is relatively low, and in
models of electroweak scale inflation, we find $\dot{\varphi}\sim
(10^{-3} {\rm eV})^2$.
Note, however, that if the potential has a more complicated shape,
$\dot\varphi$ can be higher, such as in inverted 
hybrid models~\cite{Copeland:2001qw}.
Therefore, we treat $\sd$ as a free parameter.

The dynamics of the fields are described by 
the equations of motion
\begin{subequations}
\label{equ:motion-hybrid-01}
\begin{eqnarray}
\label{equ:motion-hybrid-01a}
\partial^2 \varphi ( t, \vec{x} ) & =&
-3H\sd(t,\vec{x})
 - g^2 \chi^2 (t, \vec{x}) \varphi ( t, \vec{x} ),
\\
\label{equ:motion-hybrid-01b}
\partial^2 \chi( t, \vec{x} ) & =&
-3H\dot{\chi}(t,\vec{x})
+
\left(m^2 - g^2  \varphi^2 ( t, \vec{x} ) \right) \chi (t, \vec{x}) 
\nonumber\\&&
- \lambda \chi^3  ( t, \vec{x} ).
\end{eqnarray}
\end{subequations}
Let us first discuss the dynamics of the system in the simplest
approximation, in which the field fluctuations are neglected altogether.
The initial conditions are given by the field values at the end of
inflation,
when the inflaton $\varphi$ has just reached 
its critical value $\varphi_c$
and is
rolling down the potential with velocity $\sd$ and the $\chi$
field is still at rest in its minimum,
\begin{eqnarray}
\label{equ:motion-hybrid-02}
\varphi (0, \vec{x} ) = \varphi_c & \quad \text{and} \quad & 
 \sd  (0, \vec{x} ) = - \sd   ,
\nonumber
\\
\chi (0, \vec{x} ) = 0 & \quad \text{and} \quad & 
\dot{\chi}  (0, \vec{x} ) = 0.
\end{eqnarray}

When $\varphi$ rolls further down,
the defect field $\chi$ is displaced
out of the symmetric phase.
As far as it reacts 
much faster than the inflaton
field, we can assume that the field 
$\chi( t, \vec{x} )$ reaches immediately the 
minimum at the given value of 
$\varphi (t, \vec{x})$ and is given by
\begin{equation}
\label{equ:phiminimum01}
\chi^2( t, \vec{x} ) = \frac { m^2 - g^2 \varphi^2 (t)}{\lambda}.
\end{equation}
Substituting $\chi( t, \vec{x} )$ in the potential
$V(\chi, \varphi)$ we obtain 
the effective potential for the inflaton field:
\begin{equation}
\label{equ:effpot-01}
V_{\textrm{eff}} (\chi(\varphi), \varphi ) 
\equiv
\frac{g^2}{2\lambda}m^2\varphi^2-\frac{g^4}{4\lambda}\varphi^4.
\end{equation}
At late times, $\varphi$ oscillates around zero, and when the amplitude
is small enough, we can write down an approximate equation of motion
\begin{equation}
\label{equ:effeqmotion-01}
\ddot{\varphi} (t) + 3 H \sd (t)
\simeq -  \omega_\varphi^2 \varphi (t),
\end{equation}
where 
$\omega_\varphi^2=m^2g^2/\lambda$.
Defining $\tilde\gamma=3H/2$, we can write  the solution
as
\begin{equation}
\label{equ:sigma-hybrid-01}
\varphi (t)          =
A e^{ - \tilde\gamma t}
\cos \left( t\sqrt{ \omega_\varphi^2 - \tilde\gamma^2 }
\right).
\end{equation}
Therefore the inflaton field shows a damped
oscillatory behaviour with respect to $t$.

\section{Instability}
\label{sect:instability}

We are interested in understanding 
the dynamics of the defect field 
just after the inflaton field has crossed
the critical point $\varphi_c$. 
We take into account the fluctuations
of the defect field.
As the mass for the defect field is negative,
 the modes at low momenta
experience a fast growth. 
We ignore the fluctuations in the inflaton field, because at early times,
as we shall mainly 
be interested in the case in
which $g^2\ll \lambda$,
the effects of the interactions between 
the fluctuations of the inflaton field
and of the defect field can be neglected and, 
even if $g^2\sim\lambda$, the homogeneous mode
of the $\varphi$ field gives the dominant contribution.
At late times, the linear approximation breaks down in any case
and the results given in Sections IIIA and IIIB are not 
valid anymore.

\subsection{Instantaneous quench}
\label{sect:quench}

Let us first discuss the instability in
the case when the mass parameter of the $\chi$ field changes
instantaneously from zero to $-m^2$. 
This case has been studied before by many authors (see, for instance,
Refs.~\cite{Boyanovsky:1997rw,%
Felder:2001hj,Garcia-Bellido:2001cb,Felder:2001kt}).
For simplicity, we ignore the expansion of the universe.

In the linear approximation, the
equation of motion for a Fourier mode $\chi(t,\vec{k})$ 
of wave number $\vec{k}$ is
\begin{equation}
\label{equ:fluct01}
\partial_0^2\chi(t, \vec{k})=(-\vec{k}^2+m^2)\chi(t, \vec{k}).
\end{equation}
Modes with $|\vec{k}|< m$ become unstable and grow
exponentially.

In a quantum field theory, Eq.~(\ref{equ:fluct01}) is valid as an
operator equation, and in a classical field theory as a field
equation.
In the quantum theory, the initial state is the vacuum, and at tree
level, it is completely described by the two-point functions of the fields,
\begin{eqnarray}
\label{equ:quantum_vacuum}
\left\langle\chi^*(\vec{k})\chi(\vec{k}')\right\rangle
&=&\frac{1}{2|\vec{k}|}(2\pi)^3\delta^3(\vec{k}-\vec{k}'),\quad
\nonumber\\
\left\langle\pi^*(\vec{k})\pi(\vec{k}')\right\rangle
&=&\frac{|\vec{k}|}{2}(2\pi)^3\delta^3(\vec{k}-\vec{k}'),
\end{eqnarray}
where $\pi=\partial_0\chi$.
It is important to note that as long as the equation of motion is
linear, the operator nature of $\chi$ only appears in these initial
conditions, and therefore the classical field theory reproduces
exactly the same time evolution, if the initial conditions are
given by a Gaussian random field with the two-point function in
Eq.~(\ref{equ:quantum_vacuum}). Thus, although we shall treat the theory
as a classical system, the results of the linear approximation are
identical to the quantum theory.

Solving Eq.~(\ref{equ:fluct01}), 
we find that the Fourier mode $\chi(t,\vec{k})$
is given by
\begin{eqnarray}
\label{equ:mode01}
\chi(t,\vec{k})&=&
\chi(0,\vec{k})\cosh\left(t\sqrt{m^2-\vec{k}^2}\right)
\nonumber\\
&&+\frac{\pi(0,\vec{k})}{\sqrt{m^2-\vec{k}^2}}
\sinh\left(t\sqrt{m^2-\vec{k}^2}\right).
\end{eqnarray}
Therefore, the variance of the fluctuations 
$\langle\chi^2\rangle(t)$ grows as
\begin{eqnarray}
\langle\chi^2\rangle(t)&=&
\int \frac{d^3k}{(2\pi)^3}\frac{d^3k'}{(2\pi)^3}
\langle\chi^*(t,\vec{k})\chi(t,\vec{k}')\rangle
\nonumber\\&&\hspace*{-1.5cm}=\langle\chi^2\rangle(0)+
\frac{1}{8\pi^2}\int_0^{m^2}dk^2
\frac{m^2}{m^2-k^2}
\sinh^2 \big(t\sqrt{m^2-k^2}\big)
\nonumber\\
&&\hspace*{-1.5cm}\sim
\frac{m}{32\pi^2t}\exp(2mt),
\label{equ:instvariance01}
\end{eqnarray}
where $\langle\chi^2\rangle(0)$ is a divergent ``vacuum'' contribution.

Another quantity we shall consider is the power spectrum $P(k)$, by which we
mean the kinetic energy density of a given Fourier mode of the field
$\chi$,
\begin{equation}
\label{equ:defPk}
P(k)(2\pi)^3\delta^3(\vec{k}-\vec{k}')
=\langle\pi^*(t,\vec{k})\pi(t,\vec{k}')\rangle.
\end{equation}
Using Eqs.~(\ref{equ:quantum_vacuum}) and (\ref{equ:mode01}),
we find
\begin{equation}
P(k)=\frac{k}{2}+\frac{m^2}{2k}\sinh^2t\sqrt{m^2-k^2},
\label{equ:instspectrum}
\end{equation}
for $k<m$.
This shows that modes with $k\lsim m$ are amplified exponentially,
or in other words, tachyonic preheating produces a power spectrum with
an effective cutoff scale $k_*\approx m$~\cite{Felder:2001hj,Felder:2001kt}.

\subsection{Hybrid model}

The assumption that the mass parameter changes
instantaneously makes analytical calculations easy, but
it is not a very good approximation for the instability at the end of
hybrid inflation. During inflation, the inflaton field
$\varphi$ is rolling down the potential slowly, and because it is only
weakly coupled to the $\chi$ field, the effective mass parameter
$m_{\chi,{\rm eff}}^2(\varphi)$ changes typically very slowly.
This case has been studied in Ref.~\cite{Asaka:2001ez}
using the linear approximation, and also in
Refs.~\cite{Karra:1997it,Bowick:1998kd} in the context of defect formation.

We assume that the fluctuations of the inflaton field
$\varphi$ are negligible, which means that we can describe
it
by its homogeneous part $\varphi (t)$. 
Near the instability point, we can approximate
%
\begin{equation}
\label{equ:sigmasmallt}
 \varphi (t) \approx \varphi_c - \sd t.
\end{equation}

We are interested in the time-dependence of the
 fluctuations $\chi(t, \vec{k})$ of the
defect field, and at early times we can linearize 
Eq.~(\ref{equ:motion-hybrid-01b}),
%
\begin{eqnarray}
\partial^2_0 \chi (t, \vec{k} ) &= &
( m^2 - g^2 \varphi^2 (t) - k^2 ) \chi (t, \vec{k} )
\nonumber\\
\label{equ:fluctphi}
&\simeq &(2 m g \sd t -k^2) \chi (t, \vec{k} ).
\end{eqnarray}
The solution for $\chi (t, \vec{k} )$ is given 
in terms of Airy functions:
%
\begin{equation}
\label{equ:fluctphi03}
\chi (t, \vec{k} ) = 
c_A (\vec{k}) Ai \left( \omega t-\frac{k^2}{\omega^2} \right)
+ c_B (\vec{k}) Bi\left( \omega t-\frac{k^2}{\omega^2} \right),
\end{equation}
%
where we have defined $\omega=(2 m g \sd)^{1/3}$.

This shows that each mode starts growing at the time 
$t=t_k=k^2/\omega^3$, when the effective mass becomes negative
$m^2_{k, \textrm{eff}} \equiv 
- \omega^3 t + k^2 <0 $. 
We can express the coefficients $c_A(\vec{k})$ and
$c_B(\vec{k})$ in terms of the values of $\chi(\vec{k})$ and $\pi(\vec{k})$
at the time $t_k$ as
\begin{eqnarray}
c_A(\vec{k})&=&\frac{1}{2}\left[
\frac{\chi(t_k,\vec{k})}{Ai(0)}
+\frac{\pi(t_k,\vec{k})}{\omega Ai'(0)}
\right],
\nonumber\\
c_B(\vec{k})&=&\frac{1}{2\sqrt{3}}\left[
\frac{\chi(t_k,\vec{k})}{Ai(0)}
-\frac{\pi(t_k,\vec{k})}{\omega Ai'(0)}
\right],
\label{equ:cacb}
\end{eqnarray}
where $Ai(0)=3^{-2/3}\Gamma(2/3)^{-1}\approx 0.355$ 
and $Ai'(0)=-3^{-1/3}\Gamma(1/3)^{-1}\approx -0.259$.

In the very early time regime, each Fourier mode behaves as
\begin{equation}
\chi(t,\vec{k})=\chi(0,\vec{k})+
(t-t_k)\pi(0,\vec{k})+{\cal O}\left((t-t_k)^3\right),
\end{equation}
which implies
\begin{eqnarray}
\label{equ:hybvariance_gen}
\langle\chi^2\rangle(t) \!&\!-\!&\!
\langle\chi^2\rangle(0) \approx
\int_0^{\sqrt{\omega^3t}}\frac{d^Dk}{(2\pi)^D}
(t-t_k)^2\frac{k}{2}
\nonumber\\
&\!\approx\!&\!
4C_D
\frac{t^2(\omega^3t)^{\frac{D+1}{2}}}
{(D+1)(D+3)(D+5)},
\end{eqnarray}
where $D$ is the dimensionality of space and 
$C_D=2(4\pi)^{-D/2}\Gamma(D/2)^{-1}$.
In the physical case $D=3$, 
\begin{equation}
\label{equ:hybvariance02}
\langle\chi^2\rangle(t) -
\langle\chi^2\rangle(0)\approx \frac{\omega^6t^4}{96\pi^2}.
\end{equation}
The spinodal value of the
field $\chi$, on the other hand, grows linearly,
\begin{equation}
\label{equ:hybspin01}
\chi^2_{\textrm{spinodal}} ( t ) \equiv 
\frac{ m^2_{ \chi, \textrm{eff}}}{3 \lambda}
 \approx \frac{\omega^3t}{3 \lambda},
\end{equation}
and therefore at early times,
\begin{equation}
\label{equ:hybratio}
\langle\chi^2\rangle(t) -
\langle\chi^2\rangle(0)\ll \chi^2_{\textrm{spinodal}} ( t ).
\end{equation}
%
This result shows that no matter how small $\sd$ is,
there is always a period after the instability, during which the
linear approximation is valid. 
Eqs.~(\ref{equ:hybvariance02}) and 
(\ref{equ:hybratio}) seem to imply that the back reaction sets in when
\begin{equation}
\label{equ:early_tspin_3D}
t^{\rm 3D}_{\rm spin}\approx
\left(\frac{32\pi^2}{\lambda}\right)^{1/3}\omega^{-1}, 
\end{equation}
but because Eq.~(\ref{equ:hybvariance02}) is a power series in $\omega
t$, the expansion has broken down by that time, and instead, we have
to consider the asymptotic behaviour at late times.

At late times, $Ai( \omega t - k^2/\omega^2)$
decreases exponentially while
$Bi( \omega t - k^2/\omega^2)$
shows an exponential growth. 
Therefore, we 
neglect the first term in the right hand side
of Eq.~(\ref{equ:fluctphi03}). 
The asymptotic expansion of $Bi(z)$ 
for large values of $z$ is given by
\begin{equation}
\label{eq:Biasympt}
Bi(z)\sim\frac{1}{\sqrt{\pi}}z^{-1/4}\exp
\left(\frac{2}{3}z^{3/2}\right).
\end{equation}

The variance of the defect field $\chi (t, \vec{x})$ at late times
can therefore be computed as
%
\begin{eqnarray}
\label{equ:hybvariance01}
\langle\chi^2\rangle(t)&=&
\int_0^{\sqrt{\omega^3t}} \frac{d^Dk}{(2\pi)^D}\frac{d^Dk'}{(2\pi)^D}
\left\langle\chi^*(t,\vec{k})\chi(t,\vec{k}')\right\rangle
\nonumber \\
\mbox{} & \simeq &
C_D
\int_0^{\sqrt{\omega^3t}} dkk^{D-1}
\left\langle c_B(\vec{k})^*c_B(\vec{k}) \right\rangle\nonumber\\
&&\hspace*{1.5cm}
\times
Bi(\omega(t-t_k))^2,
\end{eqnarray}
where we have ignored the constant vacuum contribution
$\langle\chi^2\rangle(0)$. 
Using Eqs.~(\ref{equ:quantum_vacuum}) and 
(\ref{equ:cacb}), we have
\begin{eqnarray}
\left\langle c_B(\vec{k})^*c_B(\vec{k}) \right\rangle&=&
\frac{1}{8k}\left[3^{\frac{1}{3}}\Gamma\left(\frac{2}{3}\right)^2
+3^{-\frac{1}{3}}\Gamma\left(\frac{1}{3}\right)^2
\frac{k^2}{\omega^2}
\right]
\nonumber\\
&\approx&\frac{1}{8k}\left(2.645+
4.976\frac{k^2}{\omega^2}
\right).
\end{eqnarray}
 
Therefore, we find
\begin{equation}
\langle\chi^2\rangle(t) \sim 
\frac{C_D}{8\pi}
\int_0^{\sqrt{\omega^3t}} d kk^{D-2}\frac{2.645}{\sqrt{\omega(t-t_k)}}
e^{\frac{4}{3}\sqrt{\omega(t-t_k)}^3},
\end{equation}
which we can expand around small $k$ to obtain the asymptotic
behaviour
\begin{eqnarray}
\label{equ:phi2growth}
\langle\chi^2\rangle(t) & \sim & 
2.645\frac{C_D}{16\pi}
\Gamma\left(\frac{D-1}{2}\right)(\omega t)^{\frac{1-3D}{4}}
\times\nonumber\\&&\times
\left(\frac{\omega^3t}{2}\right)^{\frac{D-1}{2}}
\exp\left[
\frac{4}{3}(\omega t)^{3/2}
\right].
\end{eqnarray}
In the physical case $D=3$, this simplifies to
\begin{equation}
\langle\chi^2\rangle(t) \sim \frac{2.645}{64\pi}\frac{\omega}{t}
\exp\left[
\frac{4}{3}(\omega t)^{3/2}
\right].
\end{equation}

Using Eq.~(\ref{equ:fluctphi03}), 
it is also straightforward to write down the expression
for the power spectrum $P(k)$,
\begin{eqnarray}
\label{equ:predPk}
P(k)&=&
\frac{\omega^2}{8k}
\frac{
\left(Ai'(\omega(t-t_k))+\frac{1}{\sqrt{3}}Bi'(\omega(t-t_k))\right)^2}
{Ai(0)^2}
\nonumber\\&&+
\frac{k}{8}
\frac{
\left(Ai'(\omega(t-t_k))-\frac{1}{\sqrt{3}}Bi'(\omega(t-t_k))\right)^2}
{Ai'(0)^2} 
.
\end{eqnarray}
At late times, $t\gg t_k$, this
behaves as
\begin{equation}
P(k)\approx
\frac{\sqrt{\omega t}}{24\pi
k}\left(\frac{\omega^2}{0.355^2}+\frac{k^2}{0.259^2}\right)
e^{\frac{4}{3}\left(\omega (t-t_k)\right)^{3/2}},
\end{equation}
but the high-$k$ modes for which $t<t_k$, are still in
vacuum. Therefore, we conclude that that power spectrum has an
effective cutoff: The energy density in modes with $k\lsim
\sqrt{\omega^3 t}$ is exponentially high, but modes with $k\gsim
\sqrt{\omega^3t}$ are still in vacuum.

The growth of the long-wavelength modes stops around the spinodal
time $t_{\rm spin}$,
which we can estimate
by comparing Eq.~(\ref{equ:phi2growth}) with Eq.~(\ref{equ:hybspin01}) 
and ignoring all factors of order one,
\begin{equation}
\label{equ:hybtspin01}
t_{\textrm{spin}} \approx \frac{1}{\omega}
\left(\frac{3}{4}\ln\frac{\omega^{3-D}}{\lambda}
\right)^{2/3},
\end{equation}
which agrees with Eq.~(30) of Ref.~\cite{Asaka:2001ez}.
In contrast to Eq.~(\ref{equ:early_tspin_3D}), the dependence on
$\lambda$ is only logarithmic.
In most cases, the logarithm can be ignored, and we find the simple
result $t_{\textrm{spin}}\approx \omega^{-1}$.

Assuming that the back reaction does not change the overall shape of
the power spectrum $P(k)$ significantly, we can see that the final
value of the cutoff $\hat{k}$ is
\begin{equation}
k_*\approx \sqrt{\omega^3t_{\rm spin}}\approx \omega
\left(\frac{3}{4}\ln\frac{\omega^{3-D}}{\lambda}
\right)^{1/3}.
\label{equ:esthatk}
\end{equation}
Apart from the logarithmic factor, which is usually negligible, this
result agrees with Ref.~\cite{Garcia-Bellido:1998wm}

Ignoring the logarithmic factor and
using the slow-roll condition (\ref{equ:slowroll}), we can write
Eq.~(\ref{equ:esthatk})
as
\begin{equation}
\label{equ:kstar}
k_*\approx\left(\frac{1}{5.3\times 10^{-4}}
\frac{\sqrt{3}g}{8\lambda}
\frac{m^2}{M_P^2}\right)^{1/3}m.
\end{equation}
Consequently, the instability becomes weaker
if the energy scale of inflation is low, and only
non-relativistic particles are produced.

\section{Simulations}
\label{sec:simu}
In order to study the dynamics of the instability numerically, 
we approximate the system with a classical field theory, which is believed
to be a good approximation when occupation numbers are
large~\cite{Khlebnikov:1996mc}. 
At early times, the quantum correlations may play an important
role (see, e.g.,~\cite{Cooper:1997ii,Lombardo:2001vs}), and while 
it would be interesting to study
their consequences, that is beyond the scope of this paper.

For the scalar theory in 
Eq.~(\ref{equ:hybrid_pot}), the discretization of space and time 
is straightforward: We  replace
the derivatives by finite differences. It is convenient to define the
fields $\chi$ and $\varphi$ at the lattice sites and their time
derivatives
$\pi$ and $\tau$ at half way between two time steps.
The discretized equations of motion for $\chi$ and $\pi$ are
\begin{eqnarray}
\label{equ:latticeeom}
\frac{
\pi(t+\delta t/2,\vec{x})-\pi(t-\delta t/2,\vec{x})
}{\delta t}
&=&
\Delta\chi(t,\vec{x})-\frac{\partial V(\chi,\varphi)}{\partial \chi}
,
\nonumber\\
\frac{
\chi(t+\delta t,\vec{x})- \chi(t,\vec{x})}{\delta t} &=& 
\pi(t+\delta
t/2,\vec{x}),
\end{eqnarray}
and analogous equations apply to $\varphi$ and $\tau$.
We have used here the lattice Laplacian 
\begin{equation}
\Delta f(\vec{x})=\frac{1}{\delta x^2}
\sum_i \left[f(\vec{x}+\hat{\imath})
-2f(\vec{x})+f(\vec{x}-\hat{\imath})\right],
\end{equation}
where $\hat{\imath}$ is a vector of length $\delta x$
in the direction $i$.

The lattice approach can easily be used in an expanding universe, as
well, by using conformal coordinates $d\eta=dt/a$ 
and rescaled fields in which the
expansion of the universe only appears as time-dependent mass terms
$m_\varphi^2$, $m^2$ in the potential in
Eq.~(\ref{equ:hybrid_pot}),
\begin{equation}
m^2(\eta)=m^2a^2-\frac{a''}{a},
\end{equation}
where $\eta$ is the conformal time, $m$ is the physical mass and the
double prime indicates the second derivative with respect to the
conformal time.
We assume that the universe is radiation dominated, in which case $a''=0$.

In order to approximate the quantum time evolution with the classical
equations of motion, we use initial conditions that mimic the
properties of the quantum vacuum~\cite{Khlebnikov:1996mc}.
We do this by generating a Gaussian random field
with the two-point functions in Eq.~(\ref{equ:quantum_vacuum}).
This reproduces the quantum tadpole diagram correctly and can be
interpreted as a leading-order perturbative quantum correction.
However, it has the drawback that it generates radiative
corrections. This is a major problem for the inflaton mass term
$m_\varphi^2$, because it was supposed to be extremely small, but the
radiative correction gives a contribution of the order
$\delta m_\varphi^2=O(g^2/\delta x)$ in two spatial dimensions and
$\delta m_\varphi^2=O(g^2/\delta x^2)$ in three, which is relatively
large. 
Following 
Ref.~\cite{Rajantie:2001nj}, we solve this problem by
``renormalizing'' the mass term: We use a negative non-zero mass term
$-\delta m_\varphi^2$ in the equations of motion, and choose its
value in such a way that the radiative correction cancels.
In two dimensions, we need to compute
\begin{equation}
\label{equ:counterterm}
\delta m_\varphi^2=\frac{g^2}{4\pi^2\delta x}
\int_0^\pi 
\frac{d^2w}{\sqrt{\sin^2w_x+\sin^2w_x}}
\approx 0.32\frac{g^2}{\delta x}.
\end{equation}
The corresponding three-dimensional result~\cite{Rajantie:2001nj} 
for the potential in Eq.~(\ref{equ:pot3d})
is $
\delta m_\varphi^2\approx 0.452g^2/\delta x^2$.

To study the instability and growth of the
long-wavelength $\chi$ modes after the symmetry breakdown,
we used two-dimensional simulations, because 
we could  reach larger system sizes and longer
simulation times that way.
As we shall discuss in Section~\ref{sect:3dsimu}, we also carried out
three-dimensional simulations to make sure that, although 
the details of the back reaction are dependent on the
dimensionality, the overall qualitative picture is not.

In the two-dimensional simulations, we chose $\chi$ to be
a real one-component field $\chi$.
The lattice size was $L=2016^2$, and in units in which
the mass parameter was $m^2=1$, the lattice spacing was
$\delta x=0.5$.
We used two set of couplings, $g=10^{-3}$ and $g=10^{-4}$,
with $\lambda=200g^2$ in both cases.
In order to isolate the dependence on $\sd$, we used a Minkowskian
space with $H=0$. We recall that 
 $\sd$ and $H$ are related by
 Eq.~(\ref{equ:slowroll}) but we treat $\sd$
as a free parameter.

\begin{figure}
\center
\epsfig{file=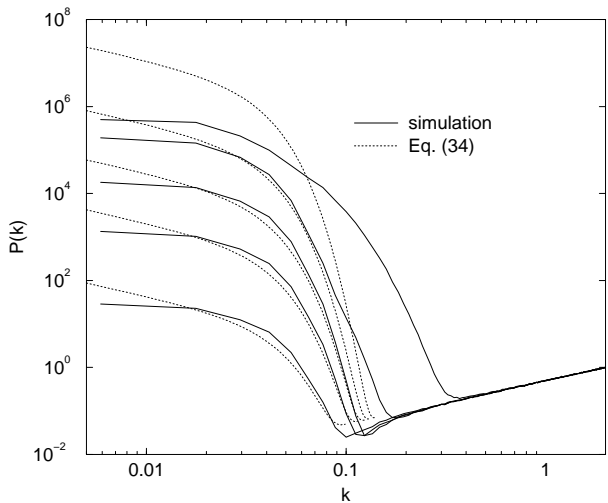,width=8cm}
\flushleft
\caption{
\label{fig:powers}
Power spectra of the $\chi$ field measured at various times after the
transition
for $g=10^{-4}$, $\sd=1.0$. 
The solid and dotted lines show the results of the numerical
simulations and of the analytical approximation in
Eq.~(\ref{equ:predPk}), respectively. 
From bottom to top, the lines correspond to the
instants when $\langle\chi^2\rangle=1$, 10, 100, 1000 and 10000.
}
\end{figure}

\begin{figure}
\center
\epsfig{file=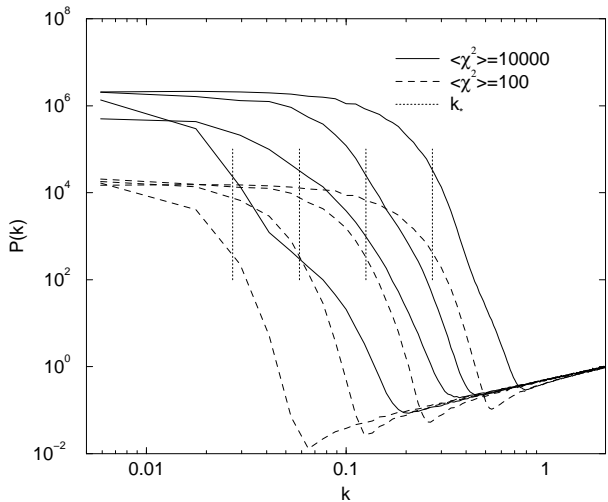,width=8cm}
\flushleft
\caption{
\label{fig:powers2}
Power spectra of the $\chi$ field at two different times 
after the transition for $g=10^{-4}$ and various different values of 
$\sd$. The dashed and solid lines corresponds to the instants when 
$\langle\chi^2\rangle=100$  and $10000$, respectively,
and the four different curves correspond to values
$\sd=0.1$, $1$, $10$ and $100$ from bottom to top.
The dotted vertical lines show the cutoff scales $k_*$ calculated 
in Eq.~(\ref{equ:esthatk}) (ignoring the logarithmic factor).
}
\end{figure}

At the start of the simulation, the inflaton field was given an
homogeneous initial value $\varphi_i =1.01\varphi_c$ and a certain
initial velocity $\sd$.
As discussed above, the initial condition for the $\chi$
field was a Gaussian random field that satisfied the ``quantum'' two-point
function, Eq.~(\ref{equ:quantum_vacuum}). We followed the time evolution of
the field configuration by solving numerically the equations of motion, 
Eq.~(\ref{equ:latticeeom}).
The perturbative mass counterterm, Eq.~(\ref{equ:counterterm}), ensured that
the inflaton $\varphi$ did not experience any significant acceleration
or deceleration before it reached $\varphi_c$.

We measured the power spectrum $P(k)$ 
of the $\chi$ field 
defined in Eq.~(\ref{equ:defPk})
at the instants when
$\langle\chi^2\rangle (t)$ was equal to some positive power of 10. 
In Fig.~\ref{fig:powers}, we compare the simulated power spectra to
the analytical result in Eq.~(\ref{equ:predPk}) for the parameter
values $g=10^{-4}$ and $\sd=1.0$. We can see that at early times, the
linear approximation works extremely well, apart from the very lowest
momenta. However, these momenta correspond to wavelengths which are of
the order of the system size, and therefore this discrepancy may well
be due to finite-size effects.
Another possibility is that the fluctuations, small as they are, give
rise to enough friction to damp down the exponential growth of the
modes with lowest $k$. Both of these effects are artifacts of our
numerical technique, and would therefore be absent in the true quantum
theory.

At later times,
when $\langle\chi^2\rangle
\approx 1000$, a discrepancy appears at high momenta, due to
the back reaction. Furthermore, when
$\langle\chi^2\rangle
\approx 10000$, we can see that 
the linear approximation has broken down.

In Fig.~\ref{fig:powers2}, we show the power spectra measured at the
instants when $\langle\chi^2\rangle
\approx 100$ and $10000$ for different values of $\sd$. At the former
time, the evolution is still linear and well described by the
analytical formula, given in Eq.~(\ref{equ:predPk}), 
while at the latter time, the
non-linearities have set it. Nevertheless, the power spectrum retains
its overall shape, with very high values at low
momenta and a sharp cutoff around the same momentum scale as in the
linear approximation. 
In particular, the back reaction does not change the
cutoff scale $k_*$, and therefore the prediction 
$k_*\approx (2mg\sd)^{1/3}$ of the
linear theory in Eq.~(\ref{equ:esthatk}) remains valid.

\section{Defect formation}
\label{sect:defects}
Depending on the number of components in 
the scalar field $\chi$, the potential
$V(\chi,\varphi)$ is invariant 
under discrete or global symmetries, which get
spontaneously broken when the inflaton $\varphi$ crosses its critical
value $\varphi_c$, if $\chi$ obtains a non-zero expectation value. 
In general, this leads to the formation of topological
defects~\cite{Kibble:1976sj}.
These defects may have several important effects on the later
evolution of the universe~\cite{VilShel}, and in many cases the fact
that they are formed can by itself rule out the model.
On the other hand, defect formation is also important, because it 
is a convenient way of probing the dynamics of the transition.

In a scalar field theory, defect formation can be understood in terms
of the Kibble mechanism~\cite{Kibble:1976sj}. (For a recent review,
see Ref.~\cite{Rajantie:2002ps}.)
When the symmetry breaks, all directions on the vacuum manifold are
equally probable, and whichever the system happens to choose is purely
a matter of chance. However  the correlation length of
the $\chi$ field cannot be infinite at the time of the transition, and
therefore the system cannot choose the same direction everywhere in
space but only inside domains whose size is determined by the
correlation length $\hat{\xi}$ at the time of the transition.

This argument was made more accurate by Zurek~\cite{Zurek:1985qw},
who related $\hat{\xi}$ to the rate of the transition and the critical
indices of the theory.
The present case in which the defects are formed at the end
of inflation is particularly simple, because, as the inflation has diluted
away all the energy density, we can assume that the
universe is at zero temperature.
Therefore the equilibrium correlation length is given by the inverse
of the mass, and near $\varphi_c$ we have simply
\begin{equation}
\label{equ:linearm2}
m^2_\chi(\varphi)=-m^2+g^2\varphi^2
\approx -2gm\sd t=-\omega^3t.
\end{equation}
Furthermore, the Lorentz invariance guarantees that the relaxation
time of the field $\chi$ is equal to the correlation length,
$\tau_\chi=\xi_\chi=m_\chi^{-1}$.

When $\varphi$ approaches $\varphi_c$, the equilibrium values of
both $\xi$ and $\tau$ diverge,
and eventually $\chi$ must fall out of equilibrium.
This happens approximately at the time when the relaxation time $\tau$
is equal to $|t|$.
Thus, we have 
\begin{equation}
\label{equ:xipred}
\hat{\xi}\approx\xi(-\tau)=(\omega^3\hat{\xi})^{-1/2}
=\omega^{-1}.
\end{equation}
This determines the typical distance between the topological defects.

Comparing Eq.~(\ref{equ:xipred}) with Eq.~(\ref{equ:esthatk}), we see
that $\hat{\xi}\approx k_*^{-1}$, and this is obviously no accident.
Indeed, a power spectrum with a cutoff $k_*$ means that the field
changes strongly on distances larger than $1/k_*$, but is smooth at
shorter distances. It is therefore clear that the correlation length
is given by $1/k_*$.
Thus, we are simply using different language to 
describe the same physical
phenomenon.

We tested the prediction in Eq.~(\ref{equ:xipred})
in the two-dimensional simulations described
in Section~\ref{sec:simu}.
Because
$\chi$ is a real one-component field, the
broken symmetry is $Z_2$ and the topological defects are
domain walls. With the parameters we were using,
Eq.~(\ref{equ:xipred}) predicts that the typical distance between
walls 
is $\hat{\xi}\approx 0.79(g\sd)^{-1/3}$. 
We can also estimate  
the total length of domain walls immediately after the phase
transition to be
\begin{equation}
\label{equ:predLwall}
L^{\rm pred}_{\rm wall}
\approx L^2/\hat{\xi}\approx 1.28\times 10^6 (g\sd)^{1/3}.
\end{equation}

\begin{figure}
\center
\epsfig{file=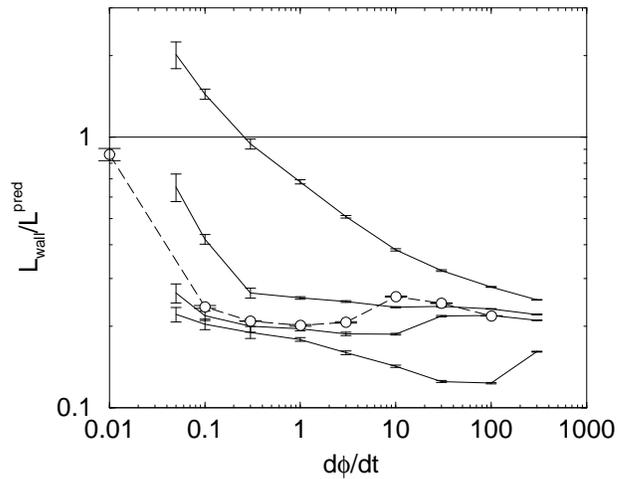,width=8cm}
\flushleft
\caption{
\label{fig:walllen}
The ratio of the measured wall length $L_{\rm wall}$ to the prediction
in Eq.~(\ref{equ:predLwall}) as a function of the velocity $\sd$. The
solid lines are for $g=10^{-4}$ and, from top to bottom, 
correspond to the times when
$\langle\chi^2\rangle=10^3$, $10^4$, $10^5$ and when
$\langle\varphi\rangle=0$.
The dashed line with open circles corresponds
to $g=10^{-3}$ and $\langle\chi^2\rangle=10^3$.
}
\end{figure}

\begin{figure}
\center
\epsfig{file=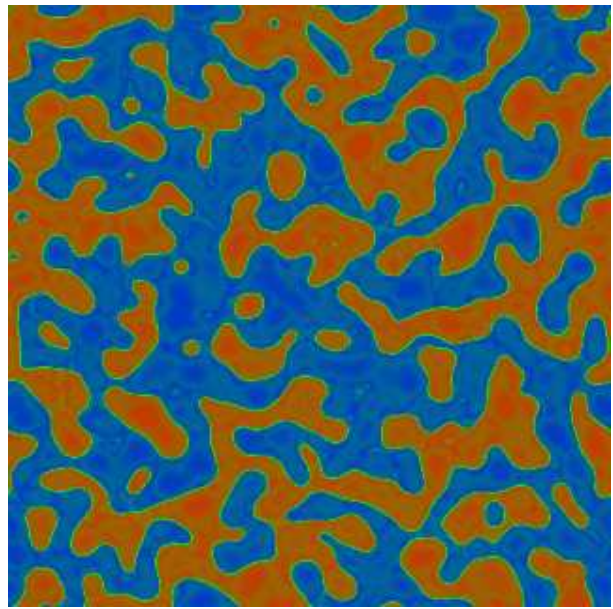,width=8cm}
\flushleft
\caption{
\label{fig:network}
An example of a network of domain walls after the transition.
Blue (dark grey) and red (medium grey) 
regions correspond to positive and negative values of $\chi$, and
$\chi$ vanishes at the 
green (light grey) 
domain walls.
}
\end{figure}

In our simulations, we measured the total length of domain walls 
$L_{\rm wall}$ in
the system by counting all pairs of neighbouring lattice sites where
$\chi$ has different signs and multiplying the result by the lattice
spacing $\delta x$.
We carried out these measurements at various times after the
transition, and the results are shown in 
Fig.~\ref{fig:walllen} for different values of $\sd$.
In the figure we plot the ratio of the measured value to
the predicted one given by
Eq.~(\ref{equ:predLwall}), and therefore a horizontal line corresponds
to a power law with the predicted exponent $1/3$.
Each data point is an average of eight different runs with different
random initial conditions, but we used the same set of eight 
initial conditions for
all values of $\sd$.

Testing the prediction in Eq.~(\ref{equ:predLwall}) is not entirely
straightforward, because, as the domain wall network evolves with
time, the measured value of $L_{\rm wall}$ depends on when the
measurement was carried out. 
This choice is more or less arbitrary, and we have used two different
ways of determining the time of the measurement: Either when
$\langle\chi^2\rangle$ crosses a certain threshold value, or when
$\langle\varphi\rangle$ crosses zero. These choices can be interpreted as
different ways of giving the quantity $L_{\rm wall}$ a precise definition.

When $\langle\chi^2\rangle$ is small, domain walls are ill-defined
objects, because the sign of $\chi$ can vary simply due to random
fluctuations. Therefore the measurements at
$\langle\chi^2\rangle=10^3$ give high values and do not
agree with the predicted power law.
Later, when $\langle\chi^2\rangle=10^4$ or 
$10^5$, the correspondence between the measurements and the
prediction (\ref{equ:predLwall}) is good at high values of $\sd$,
apart from a constant factor of around five.
Because the
prediction was intended to give the order of magnitude only, we can
conclude that the agreement is  good.

However, we can see that at very low values of $\sd$, 
the predicted power law breaks down.
The most likely explanation for this is that these extra domain walls
are generated by vacuum fluctuations, which are not strongly enough
suppressed when $\langle\chi^2\rangle$ is relatively small. This
means that the corresponding
definition of $L_{\rm wall}$ is not suitable for such slow
transitions.
Of course, we cannot rule out the
possibility that the prediction (\ref{equ:predLwall}) breaks down at
low $\sd$ in the full quantum theory, either.

We also tested how $L_{\rm wall}$ depends on the strength
of the couplings. The open circles in Fig.~\ref{fig:walllen} show 
$L_{\rm wall}/L_{\rm wall}^{\rm pred}$ 
in simulations with $g=10^{-3}$ measured at the time
when $\langle\chi^2\rangle=10^3$. This corresponds to
$\langle\chi^2\rangle=10^5$ in the simulations with $g=10^{-4}$, 
and we see that 
they agree reasonably well, and therefore the dependence on $g$  predicted
by Eq.~(\ref{equ:predLwall}) seems to be correct. Furthermore, it
seems that the velocity $\sd$ below which the prediction breaks down
is independent of $g$.

Finally, we also measured $L_{\rm wall}$ at the time when $\varphi$
reaches zero for the first time, which is typically much later than
any of the other measurements. We can see that at high $\sd$, the
power-law prediction breaks down in this case. This all illustrates
that
measuring $L_{\rm wall}$ accurately is rather difficult, because the
result depends on the time when the measurement is carried
out. Nevertheless, our results show that the prediction
(\ref{equ:predLwall}) works reasonably well in a wide range of
velocities $\sd$.

\section{Local symmetry restoration}
\label{sect:locres}

In ordinary thermal phase transitions, the defect network formed in
the transition evolves mostly dissipatively, and the domain wall loops
would simply collapse and disappear with time.
However, in the case of hybrid inflation, 
the dynamics of the inflaton field may have significant effects on the later
evolution of the system.

Eq.~(\ref{equ:sigma-hybrid-01}) shows that if the $\chi$ field were
completely homogeneous, the inflaton would oscillate with roughly the 
frequency
$\omega^2_{\varphi}\approx V''_{\rm eff}(0)=(g^2/\lambda)m^2$. However,
the presence of a defect network means that $\chi$ is far from being
homogeneous: It vanishes in the cores of the defects, and furthermore,
when the defects annihilate, they release energy, which heats up the
system locally and leads to inhomogeneities in $\chi$. In this case,
the frequency of $\varphi$ becomes dependent on the position in space.

In regions where the effective $\omega_{\varphi}$ is
lower than the average value, the $\varphi$ field will start to lag
behind, and a gradient energy starts to build up. We can study this effect
in more detail in a simple model of one scalar field with an
inhomogeneous mass term, with the equation of motion
\begin{equation}
\label{eq:bubblesmotioncomp}
\partial_0^2\varphi-\vec{\nabla}^2\varphi+M^2(\vec{x})\varphi =0,
\end{equation}
where we adopt a Gaussian shape of width $\Delta$ for $M^2$,
\begin{equation}
\label{equ:pertmass}
M^2(\vec{x})=M^2_0
\frac{\left[1-\epsilon\exp\left(-\frac{\vec{x}^2}{2\Delta^2}\right)\right]}
{1-\epsilon}.
\end{equation}
We assume that the spatial extent of the perturbation is large
compared with the mass, $\Delta\gg M^{-1}$.

We assume that initially, the field has the value
$\varphi(0,\vec{x})=\varphi$ and is at rest.
At the spatial infinity, $|\vec{x}|\rightarrow\infty$, the field
oscillates with a constant amplitude
\begin{equation}
\varphi(t,\infty)=\varphi\cos\left(\frac{M_0t}{\sqrt{1-\epsilon}}\right).
\end{equation}

\begin{figure}
\epsfig{file=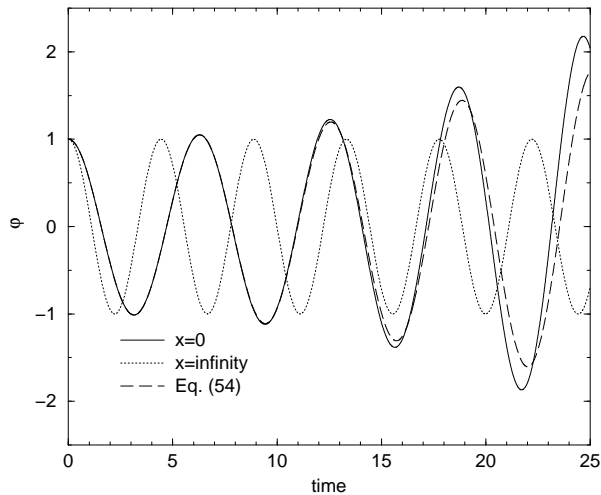,width=8cm}
\caption{
\label{fig:bubblecomp}
The amplification of the oscillations by an inhomogeneous mass term.
The dotted line shows the unperturbed oscillation, and the solid and
dashed lines show the full numerical solution and the analytical
approximation in Eq.~(\ref{equ:sol_deltasigma}) for the field at the
origin with the perturbed mass term (\ref{equ:pertmass}) and the
parameter values $D=1$, $\varphi=1$, $\epsilon=0.5$, $\Delta=10$ and 
$M_0^2=1$.
}
\end{figure}

Because $\Delta$ is large, $M(\vec{x})$ is slowly varying, and it
makes sense to treat the spatial gradient term in 
Eq.~(\ref{eq:bubblesmotioncomp}) as a small perturbation. Therefore,
we write
\begin{equation}
\varphi(t,\vec{x})=\varphi_0(t,\vec{x})+\delta\varphi(t,\vec{x}),
\end{equation}
where $\varphi_0$ satisfies the ordinary differential equation
\begin{equation}
\partial_0^2\varphi_0(t,\vec{x})+M^2(\vec{x})\varphi_0(t,\vec{x}) =0,
\end{equation}
and has the solution
\begin{equation}
\varphi_0(t,\vec{x})=\varphi\cos(M(\vec{x})t).
\end{equation}
In the equation for $\delta\varphi$, we ignore its gradient term and get 
\begin{eqnarray}
\label{equ:eom_deltasigma}
\partial_0^2\delta\varphi&+&M^2(\vec{x})\delta\varphi=\vec{\nabla}^2\varphi_0
\nonumber\\
&=&-\varphi \  t \left[
t(\vec{\nabla}M)^2 \cos (Mt)
+\nabla^2 M \sin (Mt)\right].
\end{eqnarray}
This ordinary differential equation has the solution
\begin{eqnarray}
\label{equ:gensol_deltasigma}
\delta\varphi(t,\vec{x})&=&
-\frac{\varphi \ t}{12M}
\left[
2(\vec{\nabla}\ln M)^2M^2t^2\sin (Mt)
\right.\nonumber\\&&\left.\hspace*{-.5cm}
-3\left(\vec{\nabla}^2\ln M\right)\left(
Mt\cos (Mt)-\sin (Mt)\right)
\right].
\end{eqnarray}
At the origin $\vec{x}=0$, the gradient $\vec{\nabla}\ln M$ vanishes and 
Eq.~(\ref{equ:gensol_deltasigma}) simplifies to
\begin{equation}
\label{equ:sol_deltasigma}
\delta\varphi(t,0)=\varphi
\frac{tD}{8\Delta^2 M_0}
\frac{\epsilon}{1-\epsilon}
\left[\sin (M_0t)-M_0t\cos (M_0t)
\right],
\end{equation}
where $D$ is the dimensionality of space.
This behaviour 
is shown in Fig.~\ref{fig:bubblecomp} for the parameter values
$D=1$, $\varphi=1$, $\epsilon=0.5$, $\Delta=10$ and $M_0^2=1$, together
with the full numerical solution of Eq.~(\ref{eq:bubblesmotioncomp}).
We can see that the amplitude of the oscillation is growing with time.

In the case of hybrid inflation, this means that the amplitude of the
oscillations of the $\varphi$ field can eventually exceed $\varphi_c$ at
certain localized points. Above that value, the potential becomes flat, and
$|\varphi|$ can shoot to very high values.

A somewhat 
similar phenomenon has been discussed by McDonald~\cite{McDonald:2001iv},
who showed using a linear analysis that the self-interaction of the inflaton
field may lead to the formation of ``condensate lumps''. 
In our case, the interaction with the matter field $\chi$ plays a
more important role, because the inflaton-matter coupling is stronger 
than the inflaton self-coupling
and because the instability at the end of inflation generates large
inhomogeneities in the $\chi$ field.

If the damping of the oscillations caused by the expansion of the universe or
by interactions with other fields is very weak, this overshoot may happen
already when $\varphi$ reaches its turning point for the first time.
A very rough way of estimating how likely this  is consists in calculating the
damping of the amplitude during the first oscillation, 
\begin{equation}
\label{equ:dampingest}
\frac{\varphi(t_{\rm osc})}{\varphi(0)}\sim
\exp\left(-\frac{3\pi H}{\omega_\varphi}\right)
\sim \exp\left(-\frac{3m}{gM_p}\right).
\end{equation}
Thus, if $m\ll gM_p$, this localized overshoot is likely
to happen at the first turning point.

\begin{figure}
\center
\epsfig{file=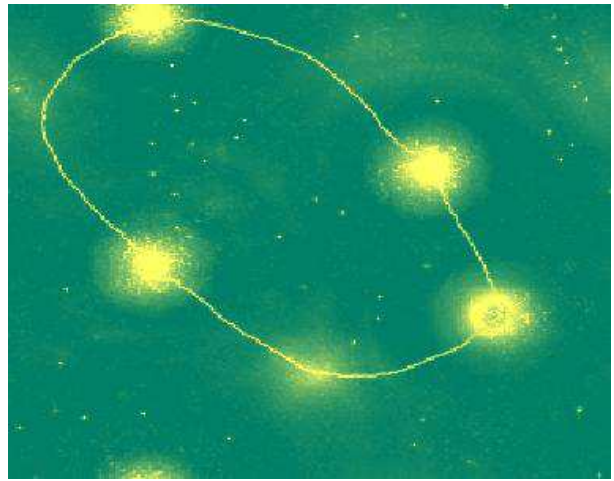,width=8cm}
\flushleft
\caption{
\label{fig:bubbles}
A snapshot of the energy density in a two-dimensional simulation a
relatively long time after the phase transition. The symmetry has been
locally restored in certain circular regions by the oscillations of
the inflaton field $\varphi$.
}
\end{figure}

Even if there is no overshoot at that time,
the amplitude of oscillations grows gradually at points where the
frequency is lower, and $|\varphi|$ may therefore exceed $\varphi_c$ at
some later time.
We can see this effect in our two-dimensional
simulations, even when we take into
account the expansion of the universe with the Hubble rate $H=0.0024$,
which corresponds to $m/gM_p\approx 0.1$.
First, the domain wall network is evolving, with cusps and loop
collapses releasing shock waves of energy. After a few oscillations of
$\varphi$, however, the energy density and the amplitude of the
oscillations of $\varphi$ in certain localized regions
start to grow. Eventually, $|\varphi|$ exceeds $\varphi_c$ in these
regions, and
the $Z_2$ symmetry is locally restored.
Because the potential of $\varphi$ is extremely flat above $\varphi_c$,
the force that pulls $\varphi$ back towards its minimum, effectively
disappears at this point.

A snapshot of the energy density at this stage is shown in
Fig.~\ref{fig:bubbles}. The bright, almost circular disks are regions
in which the energy density is very high. 
In this particular case, all these ``hot spots'' are located at a
domain wall, but this is not always the case. In some simulations,
they appeared in places where shock waves form cusps or loop collapses
hit each other.

We stress that for the existence of these hot spots, it is crucial
that in hybrid inflationary models, the inflaton potential is
extremely flat above the critical value. In a model with a convex
potential, they would only appear as regions with a slightly higher
amplitude. Even with the flat potential, the
gradient energy pulls $\varphi$ towards the origin,
and therefore it is slowly oscillating around zero with a high
amplitude. Of course, there is nothing that would stabilize these
hot spots, and therefore they eventually die away
having radiated away their energy.
It is, nevertheless, interesting to speculate whether they could have
significant cosmological consequences, because the local energy
density inside them is much higher than the reheat temperature.
In more complicated models in which the inflaton $\varphi$ is charged 
under a continuous global symmetry, the hot spots may, in fact,
become stable Q-balls~\cite{McDonald:2001iv,Kusenko:1998si}.

\section{Three-dimensional simulations}
\label{sect:3dsimu}

\subsection{Global theory}
\label{sect:3dglobal}

In order to make sure that our findings are not artifacts of the
two-dimensional theory, we also simulated a theory with three spatial
dimensions. Because domain walls are ruled out in cosmology, we used a
complex field $\chi$, which means that the topological defects
in the theory are cosmic strings. The potential of the theory is
\begin{equation}
\label{equ:pot3d}
V(\varphi,\chi)=
\frac{1}{2}m_\varphi^2\varphi^2+g^2|\chi|^2\varphi^2
+\lambda\left(|\chi|^2-v^2\right)^2,
\end{equation}
and the continuum equations of motion in conformal coordinates 
are~\cite{Rajantie:2000fd}
\begin{subequations}
\label{equ:eom3d}
\begin{eqnarray}
\partial_0^2 \varphi & =& \partial_i\partial_i\varphi
 - 2 g^2 |\chi|^2 \varphi,
\label{equ:eom3dsigma}
\\
\partial^2_0\chi &=& \partial_i\partial_i\chi
+(2\lambda v^2a^2+g^2\varphi^2)\chi
-2\lambda|\chi|^2\chi.
\label{equ:eom3dphi}
\end{eqnarray}
\end{subequations}
The time derivatives are with respect to the conformal time $\eta$, and
we have assumed that the universe is radiation dominated so that the
scale factor behaves as $a=1+H\eta$.

The discretization of the field equations was carried out in the same
way as in Sect.~\ref{sec:simu}, with lattice size $256^3$, lattice
spacing $\delta x=1.0$ and time step $\delta t=0.05$ in units with
$m^2\equiv 2\lambda v^2=1$.

To compare with the results in Ref.~\cite{Felder:2001hj}, we carried
out a simulation with the same coupling values
$g=0.01$, $\lambda=0.01$, $m=10^{15}$GeV.
In our units, the Hubble rate becomes 
$H=0.0012$.
The initial velocity of the inflaton field was
$\sd= -0.00814$, and its initial value was such that it
reaches $\varphi_c$ at time $t=10$, i.e., $\varphi(0)=100.0814$.
As in Sect.~\ref{sec:simu}, the fields also had Gaussian ``quantum''
fluctuations in the initial configuration. 

\begin{figure}
\center
\begin{tabbing}
\epsfig{file=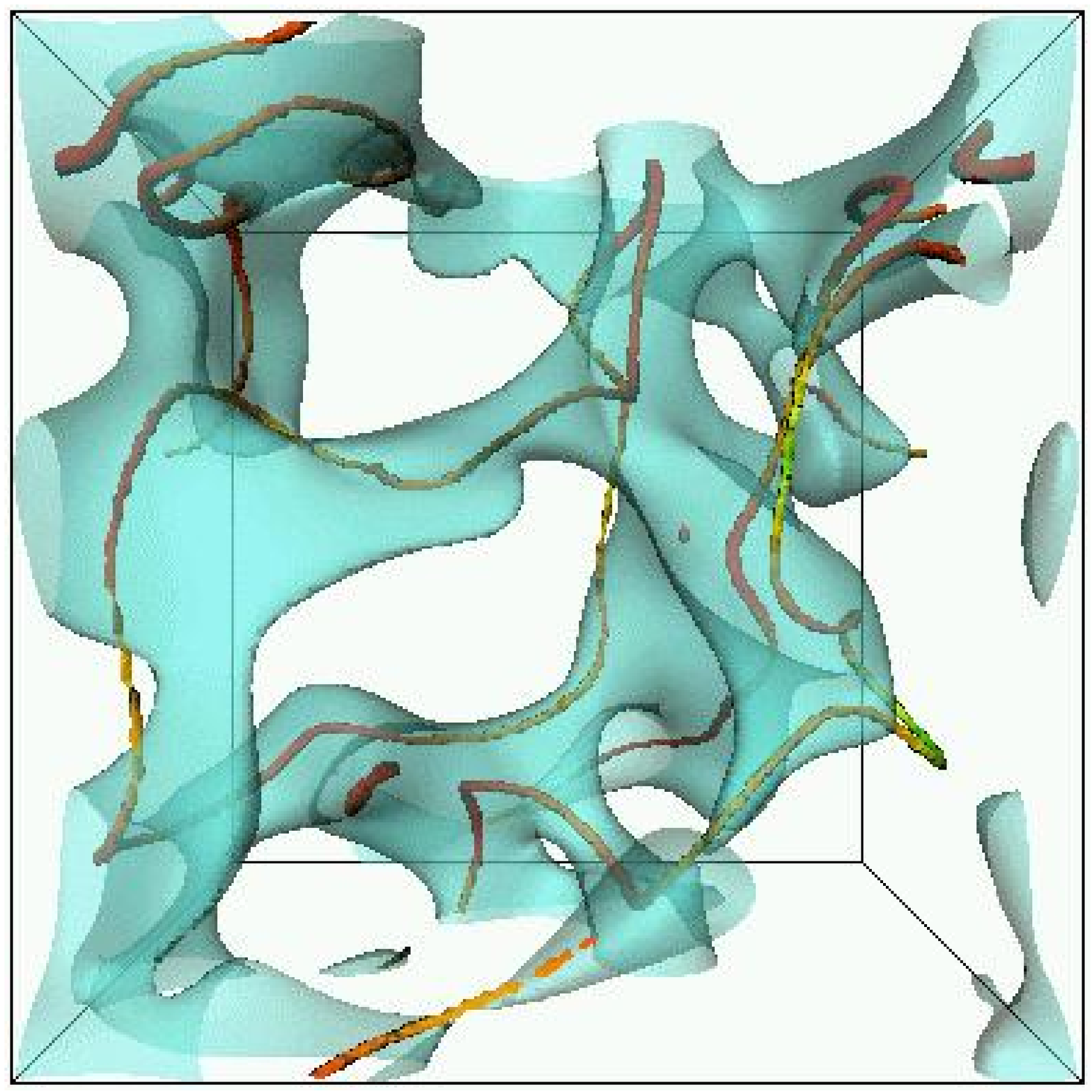,width=4cm}
\=
\epsfig{file=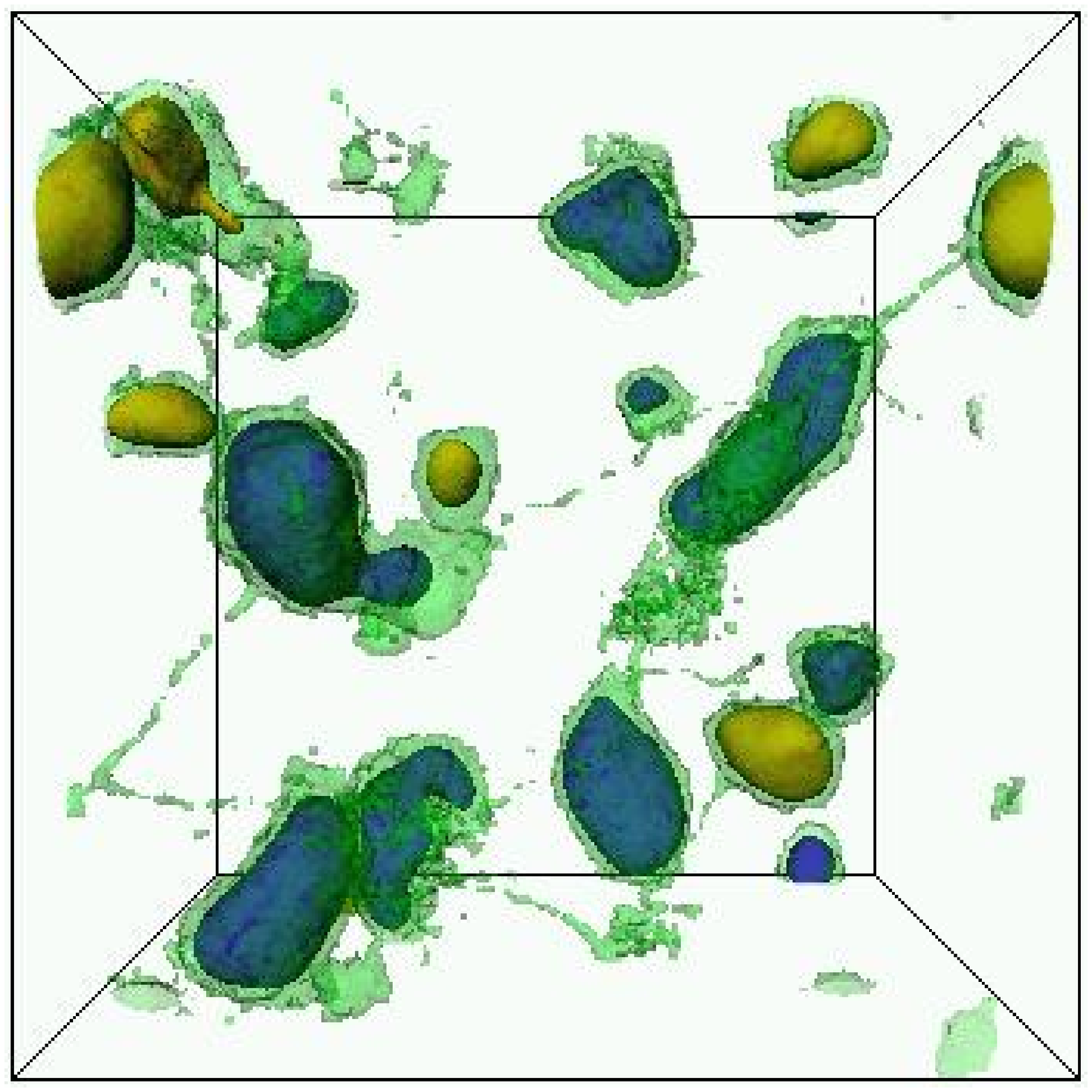,width=4cm}\\
(a)\>(b)
\end{tabbing}
\flushleft
\caption{
\label{fig:3dovershoot}
Field configurations in the global run with couplings $g=0.01$,
$\lambda=0.01$ and $m=10^{15}$GeV.\\
(a) The transparent surface shows
the isosurface $\varphi=-\varphi_c$ at time $t=224$, soon after the
spatial average of $\varphi$ reached its first turning point. The
strings correspond to the isosurface $|\chi|^2=v^2/10$ at the time
when $\varphi=0$ for the first time.\\
(b) The transparent surface shows the isosurface $|\chi|^2=v^2/10$ at
time $t=450$, and the 
blue (dark) and yellow (light) 
opaque surfaces
show the isosurfaces $\varphi=\varphi_c$ and $\varphi=-\varphi_c$ at
the same time.
}
\end{figure}

\begin{figure}
\center
\epsfig{file=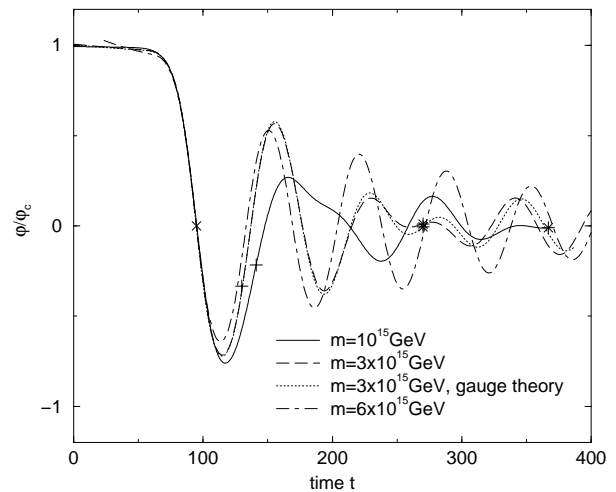,width=8cm}
\flushleft
\caption{
\label{fig:feldercomp}
Time evolution of the spatial average of the inflaton field $\varphi$.
The origin of the time axis has been shifted in each curve so that the
curves coincide when $\varphi$ first rolls down the potential.
The $\times$'s and $+$'s 
shows the times when the $\chi$ and $\varphi$
configurations were plotted in Figs.~\ref{fig:3dovershoot},
\ref{fig:3dovershoot2} and \ref{fig:3dbubblesgauge}.
}
\end{figure}

In this case
$m/gM_p\approx 0.04\ll 1$, and indeed,
the inflaton $\varphi$ overshoots $-\varphi_c$ in a significant
fraction of the space, as shown in Fig.~\ref{fig:3dovershoot}a.
We have also plotted in the same figure
an isosurface of $|\chi|^2$ at the time when
$\varphi$ crosses zero. It shows the cosmic strings formed in the
transition. The regions with $\varphi<-\varphi_c$ follow closely the
shape of the cosmic string network, in agreement with the discussion
in Sect.~\ref{sect:locres}. In these regions, the inflaton field
starts to lag behind and this effect destroys the homogeneity of the
oscillations. Indeed, Fig.~\ref{fig:feldercomp} shows that the
oscillations of the spatial average of $\varphi$ die away during the
first period, as observed earlier in Ref.~\cite{Felder:2001hj}.
In fact, a sign of this overshoot can be seen in Fig.~3 of 
Ref.~\cite{Felder:2001hj}: at $t=130$, the histogram has two peaks,
one in the broken phase, corresponding to the bulk of the space, and
one at $\varphi<-\varphi_c$.

Even though the spatial average does not oscillate,
the space is full of oscillating hot spots even at late times,
as shown in Fig.~\ref{fig:3dovershoot}b. The energy is concentrated
in these hot spots, and although
$\varphi$ oscillates with a high amplitude inside each hot spot, the
phases are uncorrelated and do not contribute to
the spatial average of $\varphi$.

\begin{figure}
\center
\begin{tabbing}
\epsfig{file=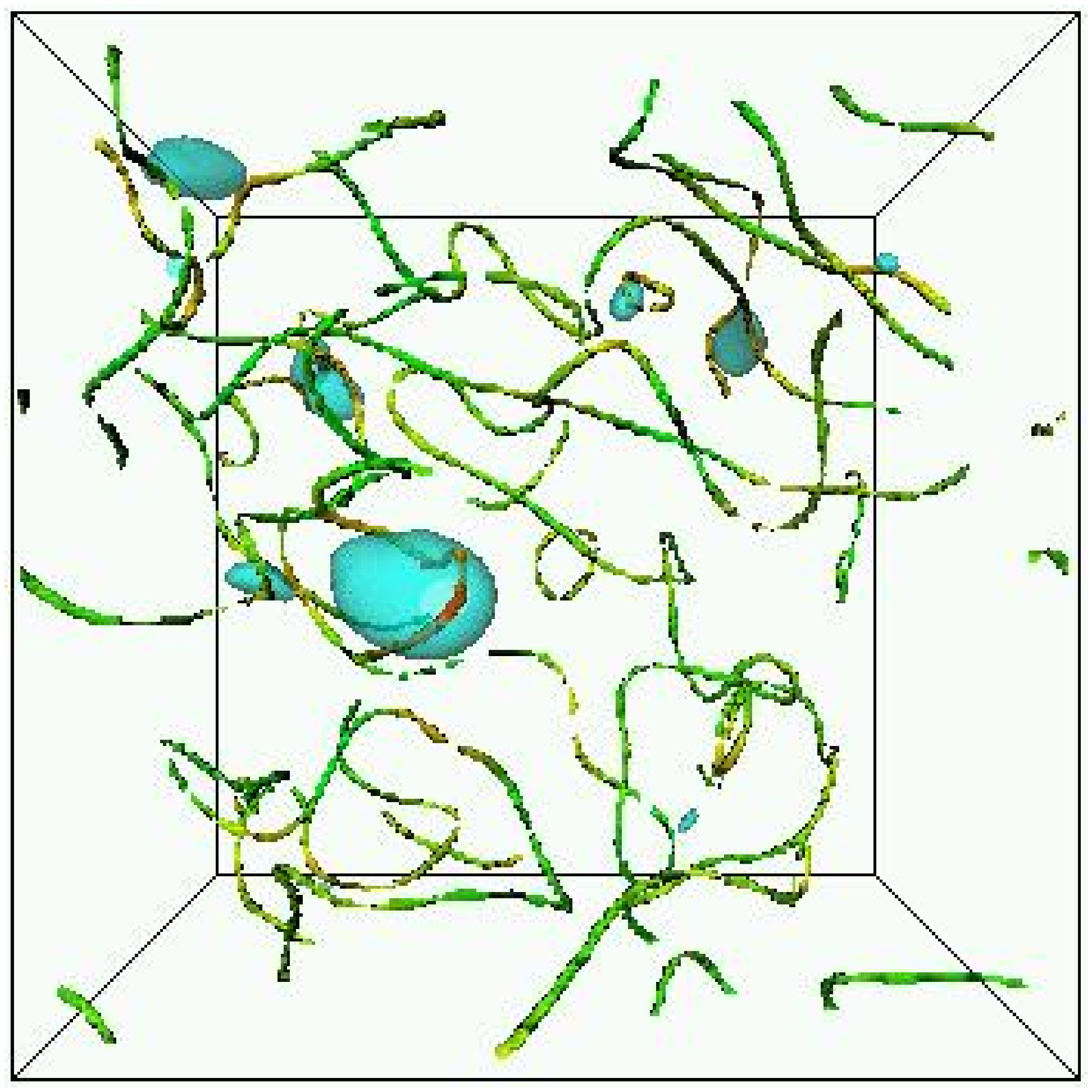,width=4cm}
\=
\epsfig{file=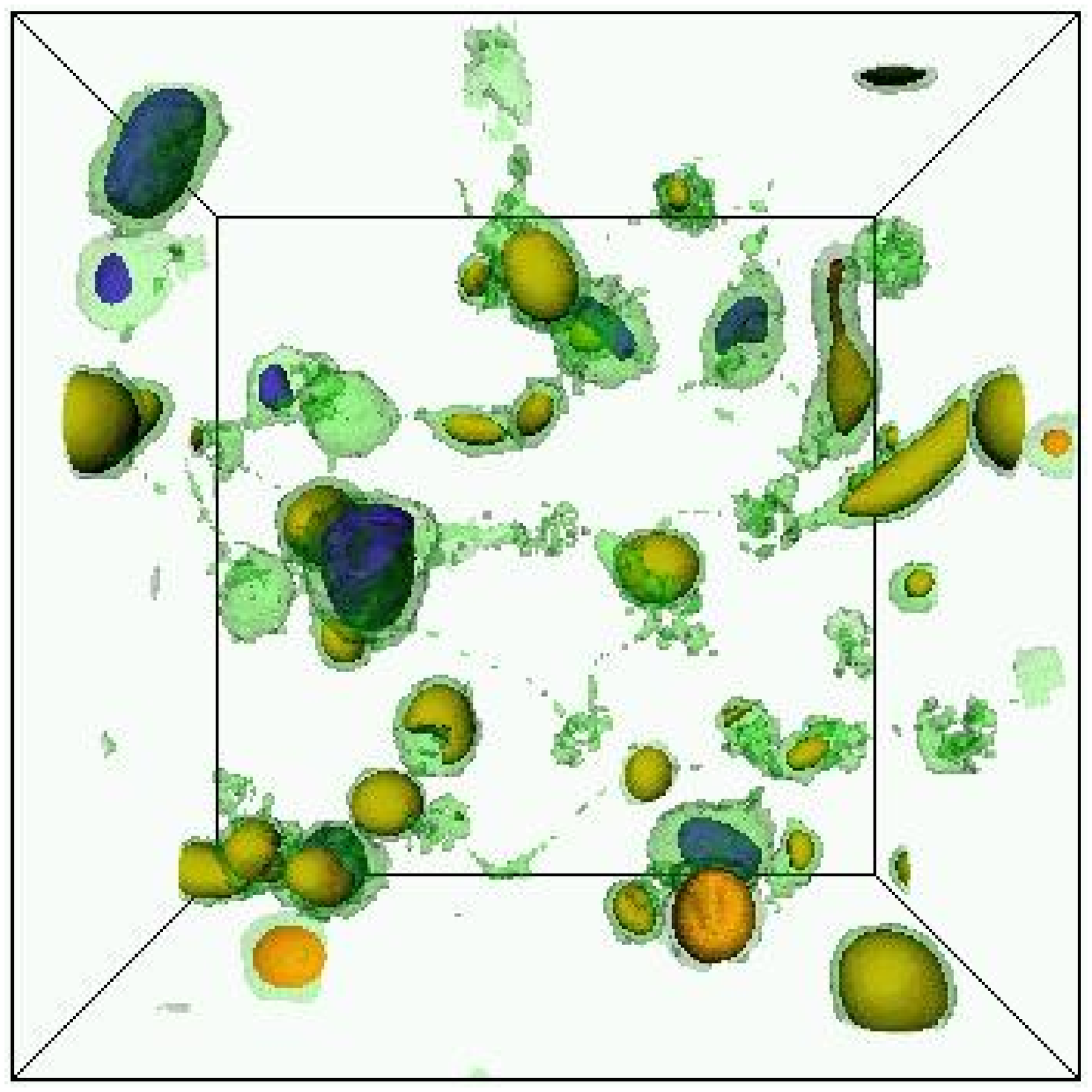,width=4cm}\\
(a)\>(b)
\end{tabbing}
\flushleft
\caption{
\label{fig:3dovershoot2}
Field configurations in the global run with couplings $g=0.01$,
$\lambda=0.01$ and $m=3\times 10^{15}$GeV.\\
(a) The transparent surface shows
the isosurface $\varphi=-\varphi_c$ at time $t=130$, soon after the
spatial average of $\varphi$ reached its first turning point. The
strings correspond to the isosurface $|\chi|^2=v^2/10$ at the time
when $\varphi=0$ for the first time.\\
(b) The transparent surface shows the isosurface $|\chi|^2=v^2/10$ at
time $t=270$, and the 
blue (dark) and yellow (light) 
opaque surfaces
show the isosurfaces $\varphi=\varphi_c$ and $\varphi=-\varphi_c$ at
the same time.
}
\end{figure}

Eq.~(\ref{equ:dampingest}) predicts that if the mass scale $m$ is
increased, the probability of an overshoot decreases, and the damping is
therefore weaker. We studied this by repeating the simulation with $m=3\times
10^{15}$~GeV, which corresponds to $H=0.00361$.
The initial values of the inflaton field and its velocity were 
$\varphi(0)=100.732$ and $\sd=-0.0732$.
For these parameters, $m/gM_p\approx 0.12$, and
Fig.~\ref{fig:3dovershoot2}a confirms that $\varphi$ exceeds
$-\varphi_c$ only in some small, isolated regions. 
The spatial average of $\varphi$ is shown
Fig.~\ref{fig:feldercomp}, and during the first oscillation, the
damping is largely due to the expansion of the universe 
[see~Eq.~(\ref{equ:sigma-hybrid-01})].

Nevertheless, as
discussed in Sect.~\ref{sect:locres}, inhomogeneities in the energy
density amplify the oscillations locally. Eventually, hot spots are
formed, and at late times, the system is again full of them, as shown
in Fig.~\ref{fig:3dovershoot2}b. This leads to the strong damping at late
times shown in Fig.~\ref{fig:feldercomp}.

In Fig.~\ref{fig:feldercomp}, we also show the time evolution of the
spatial average of $\varphi$ in a run with $m=6\times 10^{15}$~GeV,
and in agreement with the arguments in Sect.~\ref{sect:locres}, the
damping is even slower in this case. Higher $m$ corresponds to higher
Hubble rate $H$, and therefore the fact that damping becomes
weaker when $m$ increases is opposite to what the naive tree-level
treatment in Eq.~(\ref{equ:sigma-hybrid-01}) predicts.

\subsection{Gauge theory}
\label{sect:3dgauge}

\begin{figure}
\center
\epsfig{file=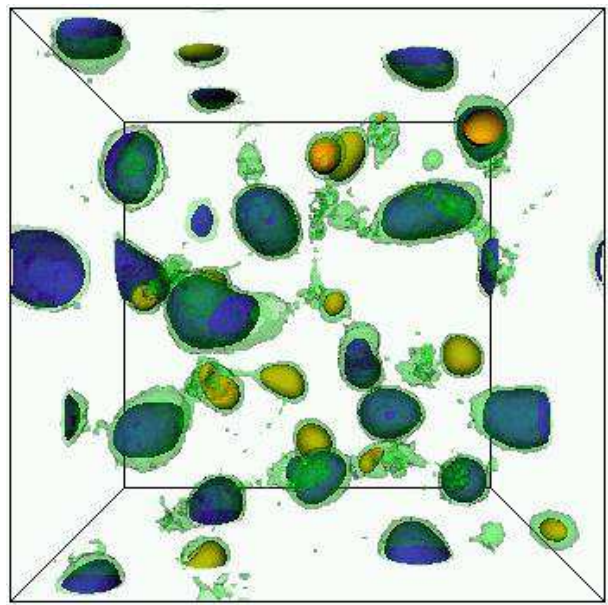,width=8cm}
\flushleft
\caption{
\label{fig:3dbubblesgauge}
A field configuration in the gauge theory with couplings $g=0.01$,
$\lambda=0.01$ and $m=3\times 10^{15}$GeV.
The transparent surface shows the isosurface $|\chi|^2=v^2/10$ at
time $t=270$, and the 
blue (dark) and yellow (light) 
opaque surfaces
show the isosurfaces $\varphi=\varphi_c$ and $\varphi=-\varphi_c$ at
the same time.}
\end{figure}

Spontaneous breakdown of a continuous global symmetry gives rise to
massless Goldstone bosons, which have not been observed in nature, and
therefore it is interesting to consider the case in which the broken
symmetry is a local gauge invariance. We need to couple the field
$\chi$ to an Abelian gauge field by replacing the derivatives in
Eq.~(\ref{equ:eom3dphi}) by covariant derivatives
$D_i=\partial_i+ieA_i$, and introducing the equations of motion for
the gauge field
\begin{subequations}
\begin{eqnarray}
\partial_0 E_i&=&\partial_jF_{ij}+2e \ {\rm Im}\chi^*D_i\chi,
\label{equ:eomgauge}\\
\partial_i E_i&=&2e \ {\rm Im}\chi^*\partial_0\chi,
\label{equ:eomgauss}
\end{eqnarray}
\end{subequations}
where 
$E_i=-\partial_0A_i$, $F_{ij}=\partial_iA_j-\partial_jA_i$.
The initial conditions for the gauge field are analogous to
Eq.~(\ref{equ:quantum_vacuum}), except that they have to satisfy the
Gauss constraint Eq.~(\ref{equ:eomgauss}).

We carried out the discretization of the gauge field equations in the
standard way as discussed, say, in Ref.~\cite{Hindmarsh:2001vp}, 
with the scalar
fields defined at the lattice sites and the gauge field on links
between the sites. 
The scalar couplings were 
$g=0.01$,
$\lambda=0.01$ and $m=3\times 10^{15}$GeV, and the gauge coupling was $e=0.1$.
For the Hubble rate, these parameters correspond to the value
$H=0.00361$.
The initial values of the inflaton field and its velocity were
$\varphi=100.732$ and $\sd=-0.0732$.

From Fig.~\ref{fig:feldercomp} we can see that the gauge field does
not change the dynamics significantly. Therefore, the existence of 
Goldstone modes
is not crucial for 
the damping.

In thermal phase transitions the gauge field plays a
crucial role in defect
formation~\cite{Hindmarsh:2000kd,Rajantie:2002ps},
but its effect depends on the temperature, and
as the transition takes place at zero temperature, the gauge fields
can be neglected. Indeed, the total length of string immediately after
the transition is similar for global and local theories.
Therefore we expect Eq.~(\ref{equ:xipred}) to be
valid in gauge field theories, as well.
Because global strings have long-range interactions while local strings
do not, the string network decays much faster in the global case.

The result that Eq.~(\ref{equ:xipred}) applies to gauge theories, 
supports the analytic estimates given in
Ref.~\cite{Copeland:2001qw} for baryon density in a scenario in which
the baryon asymmetry is generated by unwinding electroweak textures
formed by the Kibble mechanism at the end of electroweak-scale hybrid
inflation. The predicted number density of these ``knots'' is 
$\rho_H\approx \hat{\xi}^{-3}\sim k_*^3\approx 2mg\sd$.
Once formed, they can decay either by changing the topology of the
Higgs or the SU(2) gauge field configuration, and as pointed out by 
Turok and Zadrozny~\cite{Turok:1990in,Turok:1991zg}, CP
violation biases the decay of the winding configurations.
Thus, if CP violation is present, this mechanism leads to baryon
asymmetry. For a reliable calculation of the baryon density, 
one has to understand the dynamics
of the gauge field in the presence of CP violation.

The further evolution of the system confirms that 
the hot spots with local symmetry restoration
appear also in the gauge theory. Fig.~\ref{fig:3dbubblesgauge}
corresponds to Figs.~\ref{fig:3dovershoot}b and
Figs.~\ref{fig:3dovershoot2}b, and shows a very similar set of hot spots.

\section{Conclusions}
In this paper, we have studied tachyonic preheating in detail using
both analytical and numerical techniques, and taking the full dynamics
of the inflaton field into account.
We simulated the non-perturbative dynamics after the
spinodal time in two and three dimensional scalar field theories
and a three-dimensional gauge field theory, and showed that at early
times, the instability is well described by a linear approximation.
When non-linearities set in, the long-wavelength fluctuations form
topological defects, and
our results show that their number density is
related to the cutoff scale $k_*$ in the same way as in thermal phase
transitions. 

We also studied the dynamics of the system at later times,
and showed that oscillations of the inflaton field coupled
to the inhomogeneities of the matter field lead to hot spots of high
energy density, inside of which the symmetry can become temporarily
restored. Depending on the coupling constants, this effect may be very
strong, in which case it damps down the homogeneous oscillations of
the inflaton field. We believe this 
explains
\footnote{
Shortly after we released this paper as a preprint, a revised version
of Ref.~\cite{McDonald:2001iv} appeared, in which the relation between
the hot spots, or ``condensate lumps'', and tachyonic preheating was
also discussed.
} 
the high effectiveness of tachyonic
preheating observed in Ref.~\cite{Felder:2001hj}.
Further work is
required for relating the density and sizes of these hot spots to
the parameters of the model, and for understanding their other
possible consequences.

\acknowledgments
The authors would like to thank Andrei Linde, Gary Felder  
and Juan Garcia-Bellido
for useful discussions and correspondence. 
SP would like to thank L. Covi, M. Peloso 
and L. Sorbo for useful discussions.
AR was supported by PPARC.
SP was partly supported by the Marie Curie
Fellowship of the European Programme HUMAN POTENTIAL
(Contract Number HPMT-CT-2000-00096).
This work was conducted on
the SGI Origin platform using COSMOS Consortium facilities, funded
by HEFCE, PPARC and SGI.

\end{document}